\documentclass{article}
\usepackage{PRIMEarxiv}
\usepackage{soul}
\usepackage{array}

\usepackage[caption=false]{subfig}
\usepackage[table]{xcolor}

\DeclareMathAlphabet{\mathcal}{OMS}{cmsy}{m}{n}

\usepackage{graphicx} % For \resizebox
\usepackage{array}    % For centering columns

\usepackage{subfig} 
\usepackage{float}
\usepackage{caption}
\usepackage{subcaption}
\usepackage{hyperref} 

\usepackage{tabularx}   % for X columns
\usepackage{array}      % for >{\raggedright\arraybackslash}
\usepackage{booktabs}   % nicer rules (optional)
\usepackage{url}
\title{Small Cues, Big Differences: Evaluating Interaction and  Presentation for Annotation Retrieval in AR}

\author{
  Zahra Borhani \\
  Colorado State University \\
  \texttt{zborhani@colostate.edu} \\
   \And
  Ali Ebrahimpour-Boroojeny \\
  University of Illinois Urbana-Champaign \\
  \texttt{ae20@illinois.edu}
    \And
  Francisco R. Ortega \\
  Colorado State University \\
  \texttt{fortega@colostate.edu}
}

\begin{document}

%%
%% The "title" command has an optional parameter,
%% allowing the author to define a "short title" to be used in page headers.

\maketitle

%%
%% The "author" command and its associated commands are used to define
%% the authors and their affiliations.
%% Of note is the shared affiliation of the first two authors, and the
%% "authornote" and "authornotemark" commands
%% used to denote shared contribution to the research.

%%
%% The abstract is a short summary of the work to be presented in the
%% article.
\begin{abstract}

Augmented Reality (AR) enables intuitive interaction with virtual annotations overlaid on the real world, supporting a wide range of applications such as remote assistance, education, and industrial training. However, as the number of heterogeneous annotations increases, their efficient retrieval remains an open challenge in 3D environments. This paper examines how interaction modalities and presentation designs affect user performance, workload, fatigue, and preference in AR annotation retrieval. In two user studies, we compare eye-gaze versus hand-ray hovering and evaluate four presentation methods: Opacity-based, Scale-based, Nothing-based, and Marker-based. Results show that eye-gaze was favored over hand-ray by users, despite leading to significantly higher unintentional activations. Among the presentation methods, Scale-based presentation reduces workload and task completion time while aligning with user preferences. Our findings offer empirical insights into the effectiveness of different annotation presentation methods, leading to design recommendations for building more efficient and user-friendly AR annotation review systems.
\end{abstract}

%% Keywords. The author(s) should pick words that accurately describe
%% the work being presented. Separate the keywords with commas.
%% A "teaser" image appears between the author and affiliation
%% information and the body of the document, and typically spans the
%% page..

%\maketitle

\section{Introduction}

\begin{figure}[h]
    \centering
    \includegraphics[width=0.9\linewidth]{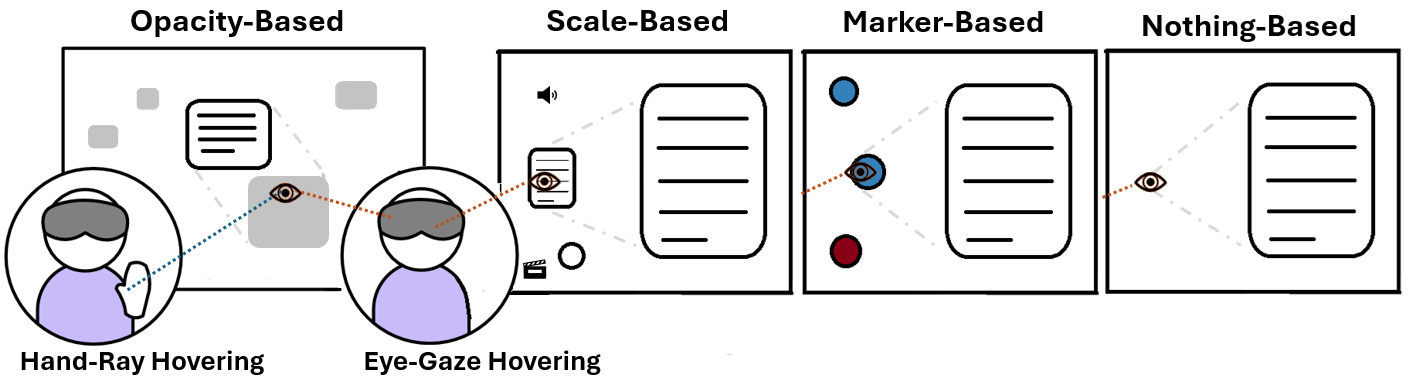}
    \caption{Comparing interaction modalities and annotation presentation methods for efficient AR annotation retrieval. Study 1 evaluates eye-gaze hovering versus hand-ray hovering interaction modalities for annotation retrieval. Study 2 examines four presentation methods (opacity, scale, marker and nothing-based) to address visual clutter.}
    \label{fig:teaser}
\end{figure}

Augmented Reality (AR) enables users to intuitively interact with virtual data and annotations overlaid on the real world, enhancing applications such as remote assistance~\cite{mohr2020mixed, zillner2018augmented, sun2016optobridge}, education~\cite{hoang2017augmented}, navigation~\cite{skinner2018indirect}, tourism~\cite{zhang2019enhancing}, medical~\cite{lin2018first, weibel2020artemis, luo2021exploring}, and industrial training~\cite{gauglitz2014world, rebol2021remote, tomlein2018augmented}. An annotation system typically consists of two components: the editor side, where annotations are created, and the reviewer side, where they are consumed or reviewed~\cite{ren2017chartaccent, 814d07b320aa437aa66dfc09d1fac612, 9757653}. Most prior research on AR annotation systems has predominantly focused on the editor side, with limited attention given to the design and usability of the reviewer experience. However, a taxonomy of annotations in outdoor augmented reality by Wither et al.\cite{wither2009annotation} and a survey on XR annotation systems by Borhani et al.\cite{borhaniSurvey} found that in most XR-HMD applications, the primary form of interaction with annotations was retrieval, not creation. Annotation creation often depended on content generated offline, either by remote collaborators using non-XR systems or through system-generated processes. Although retrieval of annotations is a common need in XR, there remains a lack of research exploring efficient methods for annotation presentation and interaction modalities to enhance the experience of the reviewer~\cite{lin2018first, yamada2018evaluation, weibel2020artemis, bai2020user}. 

Virtual annotations in prior work have often lacked mechanisms for retrieving subsets of the data, instead displaying all annotations simultaneously in the environment~\cite{marques2021remote}. However, presenting all annotations simultaneously can lead to visual clutter, making it difficult for users to focus on the annotations that are most relevant. This issue becomes especially pronounced when annotations are heterogeneous—differing in forms (e.g., text and video) and appear together in the same view. Such scenarios can quickly overwhelm users, increasing cognitive load and negatively affecting usability. These challenges underscore the importance of designing effective interaction modalities for retrieving annotations temporarily depending on the task and specific needs and presenting them in a way that is both efficient and minimally intrusive, reducing visual clutter and cognitive load for the user. 

This research aimed to investigate two key aspects of annotation review in AR: the interaction modality for retrieving annotations and the annotation presentation method for retrieval. Our first research question (RQ1) focused on comparing hand-ray hovering and eye-gaze hovering to determine which modality resulted in better performance (measured by task completion time and activation count) and user preference (in terms of fatigue and workload). Our second research question (RQ2) investigated which type of annotation presentation participants would prefer for annotation retrieval in terms of user performance and preference. We investigated these two research question in two separate studies.
 
In our first study, we compared two interaction modalities for annotation retrieval: virtual hand-ray hovering and eye-gaze hovering. Both techniques are well-suited for AR-based review systems by: (1) allowing users to temporarily access information, when needed; and (2) automatic deactivation of the augmented data, when user is no longer focusing on that annotation. User preferences for interaction modalities may differ based on the application and environment. While hand interactions are common for reviewing information and annotations in 2D environments, preferences can shift in XR environments~\cite{chiang2018augmented}. Similarly, although in AR 3D, hand interactions (e.g., VR controllers, hand rays) are often preferred for \emph{creating} annotations~\cite{blattgerste2018advantages, Flotyski2019AnnotationBasedDO}, the preferences might vary for \emph{review} systems.

%Augmented Reality (AR) uses filtering techniques to enhance user experience across various applications. These techniques can be applied to both auditory~\cite{ranjan2015natural} and visual aspects of AR to create a more natural and immersive experience.  Filtering techniques play a crucial role in refining sensor data and visual outputs in AR systems~\cite{cleaver2020rain}. Filtering techniques enhance the interaction between users and virtual environments by offering ways to selectively focus on specific areas of interest~\cite{looser2004through}.

The second study investigated four annotation presentation techniques designed to enhance visibility while minimizing visual clutter and cognitive load. The evaluated methods include Opacity-based, Scale-based, Marker-based, and Nothing-based (i.e., no visual clue) approaches (see \S~\ref{sec:methods} for details). In each condition, annotations are revealed through eye-gaze interactions. By examining user performance, cognitive workload, and user preferences across these techniques, we offer insights into their effectiveness for optimizing annotation review systems in AR environments.

% need to be fixed:

Our research provides several key contributions to XR interaction design and annotation management:

\paragraph{\textbf{Eye Gaze vs. Hand Ray.}} To the best of our knowledge, we conducted the first empirical study comparing eye-gaze hovering and hand-ray hovering (with dwell-time) interaction techniques for annotation retrieval tasks in AR. Prior works have used either of these interaction modalities as a design choice without focusing on the effect on users' performance and preference. A previous study that conducted similar comparisons did not incorporate dwell time; instead, they used a multimodal interaction that required participants to manually select the virtual data using a gesture for a selection and assembly task~\cite{tadeja2024using}. This approach differs from ours, which emphasizes dwell-based interaction for annotation retrieval. By collecting data from multiple participants and analyzing their interactions and preferences (see \S~\ref{subsec:res_study1}), we provided recommendations for designing future annotation systems, which leads to enhancing user satisfaction and efficiency.

\paragraph{\textbf{Annotation Presentation Methods.}} We evaluated four annotation presentation designs to determine their effectiveness on enhancing annotation readability and reducing user workload. To the best of our knowledge, this is the first such comparison among the annotation designs. Through empirical analysis, we showed that the choice of presentation schemes can significantly affect the user performance and preferences (see \S~\ref{subsec:res_study2}). This helped us to derive certain design implications for information retrieval in future XR systems (see \S~\ref{subsec:design_implication}).

%in terms of user performance, fatigue, and workload,

\section{Related Work}

A key aspect of data retrieval or interacting with virtual objects in AR is enabling users to achieve high precision and efficiency while minimizing cognitive workload and physical fatigue while performing tasks in visual space~\cite{ji2024research, deffeyes2011mobile}. This section reviews existing literature on interaction methods used for data retrieval, selection, and manipulation examining their strengths, limitations, and relevant findings. % on user preferences. 

\begin{table*}[t]
  \centering
  \caption{Previous interaction comparison studies. Legend: gesture (G), Eye gaze (E), Head gaze (H), and Speech (S). The bolded interaction modality represents the one that outperformed the other for the listed measurements.}
  \label{tab:previous-works}

  % global spacing just for this table
  {\renewcommand{\arraystretch}{1.2}%
   \setlength{\tabcolsep}{6pt}%

   % column layout: Year c, Ref c, big X for modalities, Tech c, two X columns
   \begin{tabularx}{\textwidth}{@{} c c >{\raggedright\arraybackslash}X c >{\raggedright\arraybackslash}X >{\raggedright\arraybackslash}X @{}}
     \toprule
     \textbf{Year} & \textbf{Ref} & \textbf{Interaction Modality} & \textbf{Technology} & \textbf{Measurement} & \textbf{Tasks} \\
     \midrule
     2025 & Ours & hand ray, \textbf{eye gaze} (dwell time) & AR-HMD & fatigue, speed & annotation retrieval \\
     2025 & \cite{lee2025experimental} & hand ray, finger point, \textbf{head gaze} & AR-HMD & accuracy, workload & pointing, tracing, rotation \\
     2024 & \cite{tadeja2024using} & hand ray+pinch, \textbf{eye gaze+pinch} & AR-HMD & speed & part selection, assembly \\
     2022 & \cite{lazaro2022multimodal} & speech, \textbf{gesture}, speech+gesture & AR-HMD & performance & identification tasks \\
     2021 & \cite{masopust2021comparison} & eye gaze, motion control & VR-HMD & fatigue & selection \\
     2020 & \cite{wang2020comparing} & \textbf{G+E+S}, G, E, G+S, E+S & AR-HMD & accuracy & selection, sliding, pressing \\
     2020 & \cite{pathmanathan2020eye} & \textbf{E+cursor}, E, H+clicker, H+gesture & AR-HMD & subjective rating & scaling, rotation, translation \\
     2018 & \cite{blattgerste2018advantages} & head gaze, \textbf{eye gaze} & VR-HMD & speed, task load, preference & selection \\
     2018 & \cite{whitlock2018interacting} & voice, \textbf{gesture}, handheld remote & AR-HMD & user preference & selection, rotation, translation \\
     2018 & \cite{afkari2018command} & \textbf{eye gaze}, physical device & 2D display & speed & command selection \\
     2017 & \cite{whitlock2018interacting} & gesture, speech, \textbf{gesture+speech} & AR-HMD & accuracy & command-based tasks \\
     2010 & \cite{kammerer2010gaze} & eye gaze, \textbf{mouse} & 2D display & accuracy, speed & web search tasks \\
     \bottomrule
   \end{tabularx}%
  }
\end{table*}

\subsection{Interaction Modalities}

Various interaction modalities have been employed in XR to retrieve or manipulate annotations, including touch gestures~\cite{tomlein2018augmented}, pen~\cite{lee2012assistive}, joystick~\cite{schmalstieg2008virtual}, hand controllers~\cite{keshav2023interaction}, touchless modalities, and wearable inputs. Touchless modalities, has had various forms in prior works, such as voice or speech commands~\cite{ryskeldiev2018spotility}, tracking sensors~\cite{garcia2020collaborative}, virtual ray~\cite{pick2016design}, head gaze~\cite{blattgerste2018advantages}, and eye gaze~\cite{mcnamara2016mobile}. Diverse techniques have also been used for wearable inputs, such as smart gloves~\cite{hsieh2016designing}, smart belts~\cite{dobbelstein2015belt}, and wrist bands~\cite{hu2020fingertrak}. All these types of interaction modalities have been used to improve user experience and efficiency which can play a vital role in adoption of these technologies. As AR devices increasingly prioritize hands-free interaction, modalities such as speech, eye gaze, and hand ray interaction have gained more attention. While voice input offers convenience, it is often unreliable in noisy environments, limiting its use in industrial or public settings~\cite{ren2024eye}. Virtual ray is a widely used interaction modality for selection tasks in AR/VR systems, particularly favored for tasks that require precise control~\cite{Dechichi2017}, such as selecting small or closely spaced targets~\cite{microsoftMixedReality}. Reitmayr et al.~\cite{reitmayr2003data} utilized a virtual ray for selecting specific topics and retrieving annotations linked to location-referenced icons within their AR-HMD interface.

Hand-ray also presents some challenges. When aiming at small targets, users often struggle with the natural jitter of hand movements, which may lead to difficulties in precise interactions~\cite{Ultraleap}. On the other hand, eye is the fastest reacting muscle in the human body, allowing for quick selections with minimal physical effort~\cite{microsoftMixedReality}. Eye-gaze interaction enables users to select and manipulate virtual content simply by looking at it, reducing the need for hand-based input~\cite{ren2016evaluating}. Furthermore, eye gaze input is often perceived by users as a form of ``mind reading,'' because the system reacts to implicit cues about what the user intends to engage with~\cite{microsoftMixedReality}. Prior studies have demonstrated that gaze-based interactions can be faster than hand-based selections for certain tasks~\cite{afkari2018command, kammerer2010gaze}, particularly when used in combination with other modalities such as voice commands~\cite{gemicioglu2023gaze}.

Tadeja et al. investigated the use of AR to improve human-robot collaboration by comparing multi-modal eye-gaze+pinch and hand-ray+pinch  methods for part selection in manual assembly tasks. Their findings indicated that while both methods offer similar perceived usability, eye-gaze interaction significantly reduces task completion time. Another previous study investigated gaze-based interaction within a simulated image-guided medical environment, where participants used gaze input to control a 2D display. The findings revealed that gaze input enabled faster task execution compared to traditional hand-based interaction with the physical device~\cite{afkari2018command}.

Eye gaze interaction has also been shown to help reduce user fatigue. One study compared the performance and fatigue progression of eye gaze interaction with traditional handheld motion controllers during continuous ten-minute VR sessions. The results revealed distinct fatigue patterns for eye-tracked input \cite{masopust2021comparison}. In another study, Blattgerste et al.~\cite{blattgerste2018advantages} demonstrated that in XR systems, eye-gaze-based aiming can be significantly faster and less fatiguing than head-gaze. Their findings showed that eye-gaze outperforms head-gaze in terms of speed, task load, required head movement, and user preference. However, gaze-based interactions may suffer from Midas Touch issues, where unintended selections occur due to involuntary gaze behavior~\cite{mohan2018dualgaze}. Hand ray interaction, on the other hand, provides a more deliberate selection mechanism, often used in AR for precise object manipulation~\cite{microsoftMixedReality}. Table~\ref{tab:previous-works} presents a summary of prior studies that compared interaction modalities for various tasks in AR environments.

\subsection{Annotation Presentation}

 Visual clutter is a common challenge in AR annotation systems, particularly in complex environments with numerous overlapping annotations~\cite{bell2001view, azuma2003evaluating}. To address this, researchers have proposed various techniques to present annotation in XR systems, including marker-based~\cite{leo2021interactive, dominic2020exploring, langlotz2013audio, wither2005pictorial}, scaling strategies~\cite{ren2024eye}, spatial grouping~\cite{garcia2020collaborative}, and adaptive content presentation~\cite{mcnamara2016mobile,azuma2003evaluating,madsen2016temporal}. 

In several previous studies, annotations were not continuously visible; instead, they appeared only when users directed their attention to specific spatial locations (without any visual marker). For example, Skinner et al.~\cite{skinner2018indirect} displayed building labels as users explored an outdoor AR city environment. McNamara et al.~\cite{mcnamara2016mobile} implemented a gaze-controlled view management technique for annotation retrieval in a grocery store setting, where product labels were automatically displayed when users looked at specific items. Although the results did not reveal statistically significant differences between the two approaches, the authors suggested that this outcome may have been due to the relatively low difficulty of the task~\cite{mcnamara2016mobile}.
 
Several previous studies have used visual markers to indicate the location of annotations embedded in the environment, with the annotation content encapsulated within these markers. Garcia et al.~\cite{garcia2020collaborative} employed color-coded spheres to encapsulate heterogeneous annotations (e.g., images, 3D models) on an AR tablet. Similarly, Leo et al.~\cite{leo2021interactive} used blue spheres within a mobile AR application to indicate the positions of note annotations added to a heart model. Langlotz et al.~\cite{langlotz2013audio} used spherical markers to visually guide users toward spatial audio annotations in a mobile augmented reality platform. In their system, green spheres indicated that an audio sticky comment was currently playing, while red spheres signified that the comment was out of the user’s focus. Similar to the two aforementioned studies that used spheres as markers, Wither~\textit{et al.} employed color-encoded spheres as depth cues to enhance users’ depth perception~\cite{wither2005pictorial}. 

In another study, Ren et al.~\cite{ren2024eye} used eye gaze assisted finger for typing in AR-HMDs and their system enlarged and highlighted recommended text input options based on the predicted results. However, in most previous studies, annotations were displayed at their actual size and in their original form within the AR or VR environment~\cite{radu2021virtual, zhang2019enhancing, mohanty2018kinesthetically, tomlein2018augmented}. While this approach may be suitable when only a few annotations are present, it becomes problematic in scenarios with a large number of annotations, heterogeneous annotations, and dense information. Displaying annotations at full size, with constant opacity and permanence, can lead to visual clutter and increased cognitive load for the user. Our research extends prior work by directly comparing the four annotation presentation methods (opacity-based, scale-based, marker-based, and nothing-based) and evaluating their effect on user efficiency and preferences.

%McNamara et al.~\cite{mcdowall1990implementation} explored a gaze-controlled view management technique for annotation retrieval in a grocery store setting, comparing it against a baseline condition with no management technique. Participants used the system to identify product names. Although the results did not reveal statistically significant differences between the two approaches, the authors suggested that this outcome may have been due to the relatively low difficulty of the task.
%The findings from these studies can inform the design of more efficient annotation review systems, contributing to improved usability in various AR applications.

%The results of these studies contribute to the growing body of research on AR interaction design by identifying efficient hands-free interaction techniques and visualization methods that can enhance annotation review tasks. Our findings have implications for AR applications in collaborative work, training simulations, and medical imaging. 

\section{Methodology}
\label{sec:methods}

This paper consisted of two studies, two tasks for Study 1 and four for Study 2. In addition, Study 1 featured two practice trials for interaction modalities and questionnaires, while Study 2 included one practice trial.  %Figure~\ref{fig:mainScene} shows a screenshot of the home scene of our experiment which includes five main tasks and two practice trials. 

%\begin{figure}[t]
%    \centering
    %\includegraphics[width=0.8\linewidth]{figures/MainScene3.jpg}
   % \caption{A screenshot of the home scene that encompasses pointers to all seven AR tasks, including the two trials (Trial 1 and Trial 2) and the five main tasks (Opacity, Ray, Scale, Marker, and Nothing).}
   % \label{fig:mainScene}
%\end{figure}

Each task was structured as a question-and-search activity, designed to evaluate user performance across three key interaction domains: visual search, content reading, and audio listening. In each task, participants were presented with five questions that required them to engage with different types of virtual annotations (text, image, audio, video, and shapes). To ensure fair comparisons, the questions were crafted to maintain a consistent difficulty level across all tasks. Some examples of these questions are as follows: ``What is the number shown inside the red circle?'' (visual search), ``What is the name of the fourth white key on the keyboard?'' (content reading), and ``What is the audio about?'' (audio listening).  Each question had a structurally equivalent counterpart in other tasks, allowing for within-subject comparisons of interaction modality and annotation presentation design. We also made sure that none of the questions needed prior knowledge. Participants were instructed to answer each question one at a time: first reading the question, then reviewing the virtual annotations, and finally submitting their answers before proceeding to the next question. Detailed information about the procedure can be found in \S~\ref{subsec:procedure}.

\begin{figure}[t]
    \centering
    \includegraphics[width=0.6\linewidth]{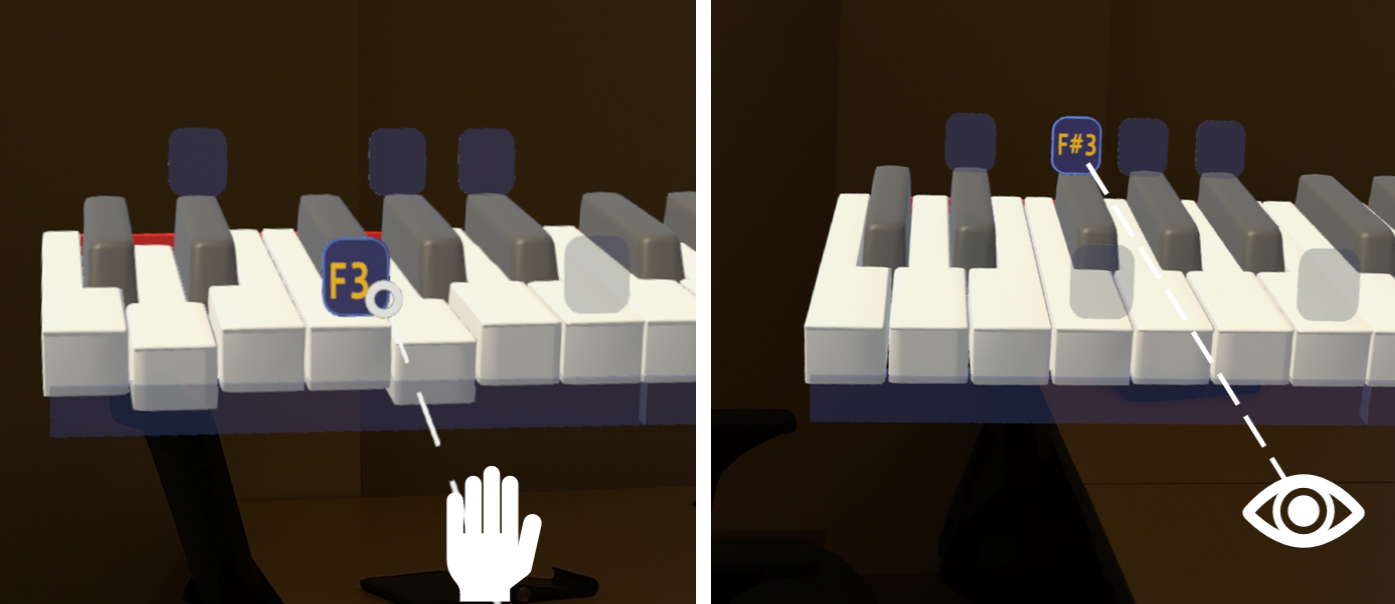}
    \caption{Two interaction modalities compared in study 1: hand-ray hovering (left) and eye-gaze hovering (right).}
    \label{fig:comparison-ray-gaze}
\end{figure}

Study 1 compares two interaction modalities, eye-gaze hovering and hand-ray hovering, for retrieving annotations to answer questions presented in a virtual questionnaire.
Each annotation was activated when the user’s gaze or hand ray hovered over an annotation for a sustained period, known as dwell time. During the hand-ray condition, participants interacted using their dominant hand, where a ray emanated from the palm to enable pointing and interaction. In this study, following the setup introduced by Blattgerste et al.~\cite{blattgerste2018advantages}, we set the dwell time threshold to 1 second of continued interaction, as it was shown to provide a responsive and natural interaction experience while minimizing false positives. Figure~\ref{fig:comparison-ray-gaze} illustrates the two interaction techniques evaluated in this study.

%\subsection{Study 1: Eye-Gaze vs. Hand-Ray Hovering}
%\label{Study1_Eye-Gaze_vs_Hand-Ray_Hovering}

%Eye-gaze hovering is an interaction modality that utilizes natural eye movements to interact with virtual objects. By leveraging eye-tracking technology, systems detect a user’s gaze and interpret it as input. This enables users to review virtual annotations simply by looking at them for a sustained period, known as dwell time. Dwell time refers to the duration a user must fixate on an object or interface element before it is activated. In this study, we used a dwell time of one second fixation to trigger the activation of virtual annotations. 

%\subsection{Study 2: Annotation Presentation Designs}
%\label{Study2_Gaze-hover_Annotation_Visualization_Methods}

%Gaze offers a natural, hands-free way of targeting content in XR and is especially useful in scenarios where users' hands are occupied. 
Study 2 examines four distinct annotation presentation methods for eye-gaze hovering to better understand user performance, preferences, and cognitive load. Figure~\ref{fig:four-annottaion-presentation} illustrates the four annotation presentation methods for the virtual piano used in the main AR tasks of this experiment: Opacity-based, Scale-based, Marker-based, and Nothing-based. The description of each method is as follows:

\begin{figure*}[h]
    \centering
    \includegraphics[width=\linewidth]{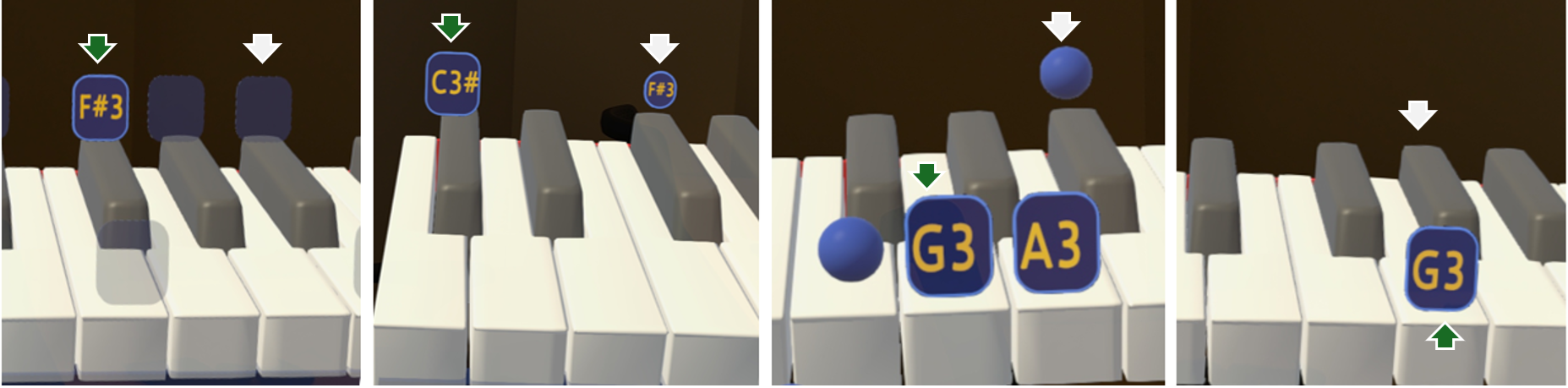}
    \caption{The four annotation presentation designs for the piano, shown from left to right, are: Opacity-based, Scale-based, Marker-based, and Nothing-based. The white arrow indicates the annotation presentation before eye gaze hovering (pre-activation), while the green arrow indicates its presentation after hovering (activated).}
    \label{fig:four-annottaion-presentation}
\end{figure*}

\begin{itemize}
    \item \textbf{Opacity-Based:} It modifies the transparency of annotations based on the gaze input. When a user looks at an annotation, the transparent shadow of annotation becomes more opaque, making it clear and readable. Conversely, when the gaze is removed, the annotation fades to maintain an uncluttered view of the environment. This approach is designed to reduce visual overload while ensuring that users remain aware of the presence of annotations at specific spatial locations. 
    
    \item \textbf{Scale-Based:} It enlarges annotations in response to gaze interaction. When an annotation is hovered over with the user's gaze, it expands to provide better visibility and readability. Once the gaze shifts away, the annotation shrinks back to its original size. This method is particularly useful when the readable size of annotations leads to overlapping with each other. By displaying compact annotations that only enlarge when necessary, this method optimizes space to reduce distractions. 
    
    \item \textbf {Marker-Based:} This method encapsulates the annotation inside a visual marker, spheres in this study, to draw attention to the location of the annotation while the information is not shown. This form of representation separates annotations from background elements, while allowing for small visual clues that avoid visual clutter and overlapping annotations. 

     \item \textbf{Nothing-Based:} In this presentation method, annotations are invisible and only appear when the user looks at a specific location or object where the annotation exists. This method completely avoids visual clutter by removing any visual clue about the annotations which might come at the cost of missing some annotations when needed.

\end{itemize}

\section{Experiment}
\label{sec:experiment}

In this section, we provide details of our experiments, which were approved by the CSU Institutional Review Board (IRB); in \S~\ref{subsec:exp-design} we explain the design of our experiments. We then elaborate on the participants (\S~\ref{subsec:exp-participants}) and equipment (\S~\ref{subsec:exp-quipment}) involved in our studies. Finally, we explain the procedure used in our experiments in \S~\ref{subsec:procedure}.

\subsection{Experimental Design}
\label{subsec:exp-design}

Study 1 employed a within-subjects design with  the interaction modality (eye-gaze hovering vs. virtual hand-ray hovering) as its independent variable. The order of interaction modalities was counterbalanced across participants to avoid biases. Each participant used both interaction modalities to retrieve annotations to answer five questions in each AR task. The hypotheses for study 1 are as follows:

\begin{itemize}
    \item \textbf{{$H_1$:} The eye-gaze hovering modality will result in shorter task completion time than the hand-ray hovering method.} This hypothesis was grounded in prior research that demonstrated the efficiency of eye gaze input over physical devices in 2D displays~\cite{afkari2018command}, as well as findings from a study comparing multimodal interactions, hand ray+pinch gestures versus eye gaze+pinch gestures for selection tasks~\cite{tadeja2024using}. %$H_1$ was not supported by our results. 
     \item \textbf{{$H_2$:} The eye-gaze hovering method in AR-HMDs will lead to more unintentional activation of the annotations than the hand-ray hovering method}. It was formed upon findings of a previous study that compared the accuracy of a traditional mouse and eye gaze~\cite{kammerer2010gaze}.
      %$H_2$ was supported by our results
        \item \textbf{{$H_3$:} The hand-ray hovering method in AR-HMDs will cause more fatigue than the eye-gaze hovering method}. This hypothesis was developed to extend the findings of a prior study that compared eye-gaze and motion controller interactions in VR in terms of fatigue.
    %$H_4$ was not supported by our analyses excepet for struggle to maintain effort item.
    \item \textbf{{$H_4$:} The eye-gaze hovering modality will result in lower cognitive load than the hand-ray hovering method.} This hypothesis was informed by previous findings from a study comparing head-gaze and hand-ray interactions~\cite{lee2025experimental}. %$H_2$ was not supported by our results.
  
\end{itemize}

Study 2 followed a 4-factor design, with annotation presentation design (Opacity-based, Scale-based, Marker-based, or Nothing-based) as the independent variable. Twenty participants performed Study 2 in a within-subject design. The hypotheses for study 2 are as follows:
 \begin{itemize}
     \item \textbf{$H_5$: There will be noticeable differences in user performance across the different annotation presentation methods.} %$H_5$ was supported by our results.
     \item  \textbf{$H_6$: There will be significant differences in user preferences across the different visual filtering methods.} %$H_6$ was supported by our results.
 \end{itemize}

\subsection{Participants}
\label{subsec:exp-participants}

~\paragraph{\textbf{Study 1.}} Fourteen participants (7 female and 7 male) took part in Study 1, ranging in age from 23 to 38 years (M = 28.64, SD = 4.3). All had normal or corrected-to-normal vision: 35\% wore glasses, none used contact lenses, and the remainder required no visual aids. No participants reported perceptual disabilities. Most participants were right-handed (86\%), with the remainder left-handed (14\%); none were ambidextrous.
~\paragraph{\textbf{Study 2.}} Twenty participants (11 female and 9 male) took part in Study 2, aged 24 to 37 years (M = 29.5, SD = 3.4). All reported normal or corrected-to-normal vision: 29.4\% wore glasses, 5.6\% used contact lenses, and the rest required no visual aids. No participants reported perceptual disabilities. Eighty percent were right-handed and 20\% left-handed, with no ambidextrous individuals.
~\paragraph{\textbf{Background and Experience}} Across both studies, all participants held graduate degrees and represented diverse fields, including engineering (civil, computer, mechanical, chemical, systems, and robotics), computer and information sciences, economics, regional planning, neuroscience, plant pathology, and music education. Regarding prior exposure to immersive technologies, 57\% of Study 1 participants (50\% in Study 2) had never used a virtual reality head-mounted display (VR-HMD). Similarly, 86\% (Study 1) and 75\% (Study 2) had never used an augmented reality head-mounted display (AR-HMD). Among those with VR experience, 36\% (Study 1) and 40\% (Study 2) reported rare use (2–3 times per year), while the rest used VR more frequently (1–2 times per month). For AR, 14\% of Study 1 participants reported rare use, and no participants in either study reported frequent AR use.

%Regarding involvement in XR research, 8.6\% had participated for less than one semester, 11.4% for less than two years, and 8.6% for three to seven years.

\begin{figure*}[h]
    \centering
    \includegraphics[width=\linewidth]{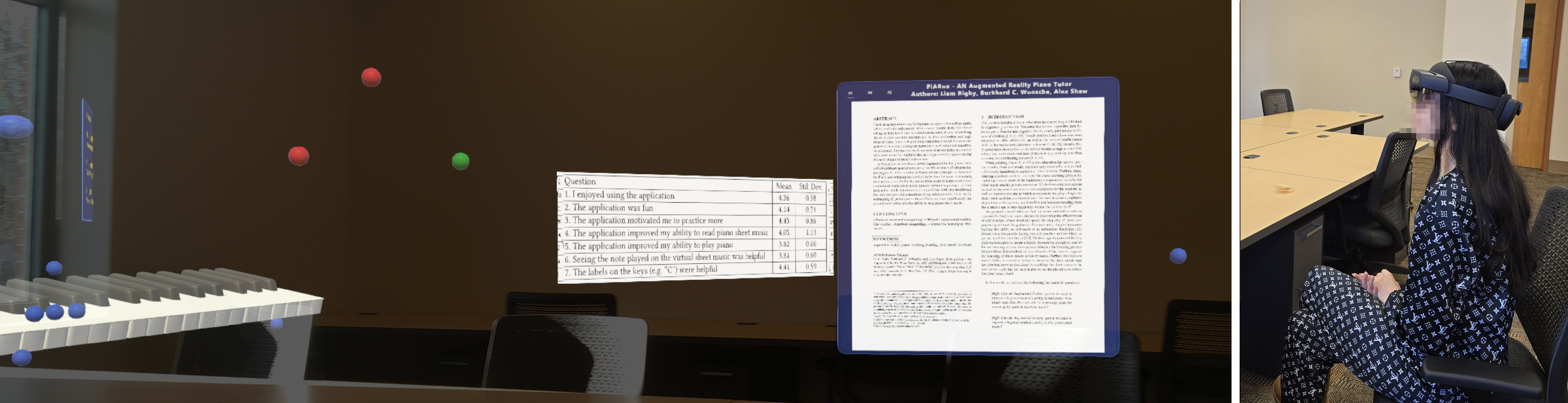}
    \caption{A participant seated and reviewing annotations (right) and a screenshot of what a participant sees in the Marker-Based task (left).}
    \label{fig:experimentTasks}
\end{figure*}

\subsection{Equipment}
\label{subsec:exp-quipment}

The 3D environment was developed using Unity 3D version 2021.3.21~\cite{Unity} and Visual Studio 2022. The AR environment was built on a computer equipped with an Intel i7-6700K CPU and an NVIDIA GeForce GTX 1080 GPU. Augmented reality features were implemented with Mixed Reality Toolkit 3 (MRTK3)~\cite{MRTK3} and deployed on a Microsoft HoloLens 2. To optimize usability, standard MRTK3 graphical components were utilized for fonts and primary user interfaces.

Audio annotations were generated using Narakeet~\cite{Narakeet}, an online tool that uses artificial intelligence to turn text into natural-sounding speech.  Similarly, video voiceovers were produced through VEED~\cite{Veed}, another online platform providing AI-based audio narration. The contents of all audio and video annotations had similar level of difficulty, with durations of 11 seconds and 13 seconds, respectively in all scenes.

\subsection{Procedure}
\label{subsec:procedure}

Participants entered the experiment room, read the consent form, and completed a pre-questionnaire. The study was conducted entirely in a seated position, with all targets displayed in front of the participant’s field of regard. The experimenter was seated behind and to the left of the participant. The experiment setup is shown in Figure~\ref{fig:experimentTasks}. The pre-questionnaire collected demographic information, participants' dominant hand, use of visual aids, and prior experience with AR/VR and computer games. To ensure data confidentiality, each participant was assigned a unique participant ID.

\begin{figure*}[h]
    \centering
    \includegraphics[width=\linewidth]{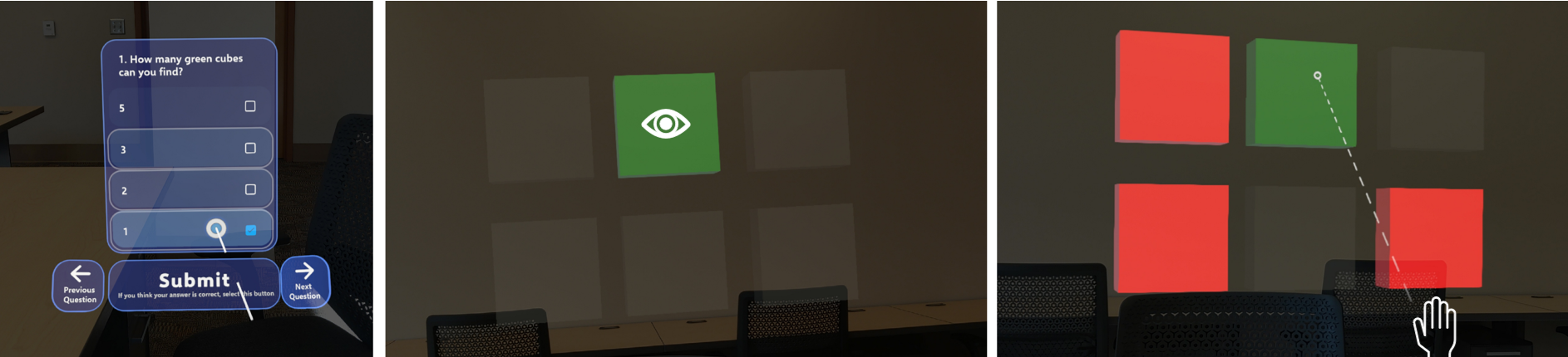}
    \caption{From left to right: (1) AR questionnaire; (2) Trial 1 containing eye-gaze hovering; and (3) Trial 2 containing hand-ray hovering.}
    \label{fig:trial}
\end{figure*}

At the start of the experiment, the experimenter provided a brief demonstration of the tasks and reminded participants that their participation was voluntary, allowing them to withdraw at any time. Before beginning the training task, participants calibrated the eye tracker. To become familiar with the environment and interaction methods, participants performed two trial tasks (eye gaze and hand-ray) in Study 1, while those in Study 2 completed only the eye-gaze trial at the beginning of the experiment. Figure~\ref{fig:trial} shows the trial environment. 

Each trial contains six transparent cubes arranged in a 2×3 grid and a questionnaire on the right side of the user. Each trial questionnaire consisted of two questions that prompted participants to count and report the number of cubes of a specified color presented in front of them (e.g.,``How many green cubes can you find?''). During the trials, transparent cubes turned into a solid color when activated by the user's gaze hover in trial 1 or hand-ray hover in trial 2. These trials were designed to provide participants with practice in activating annotations using both interaction modalities (eye gaze and hand ray) and to familiarize them with virtual questionnaires.

 Study 1 contained two main tasks: eye-gaze hovering and hand-ray hovering. Study 2 included four main tasks related to annotation appearance design. The order of tasks in both studies was randomized using a Latin square design to mitigate ordering effects. Each task comprised five questions presented through a virtual questionnaire. Participants were instructed to answer the questions sequentially by reviewing annotations. In study 1, after completing first task, participant answered the fatigue and workload questionnaires. The same procedure was followed for the second task. Upon completing the study, participants responded to a customized comparative preference and fatigue questionnaire. 

\begin{figure*}[t!]
    \centering
    \includegraphics[width=\linewidth]{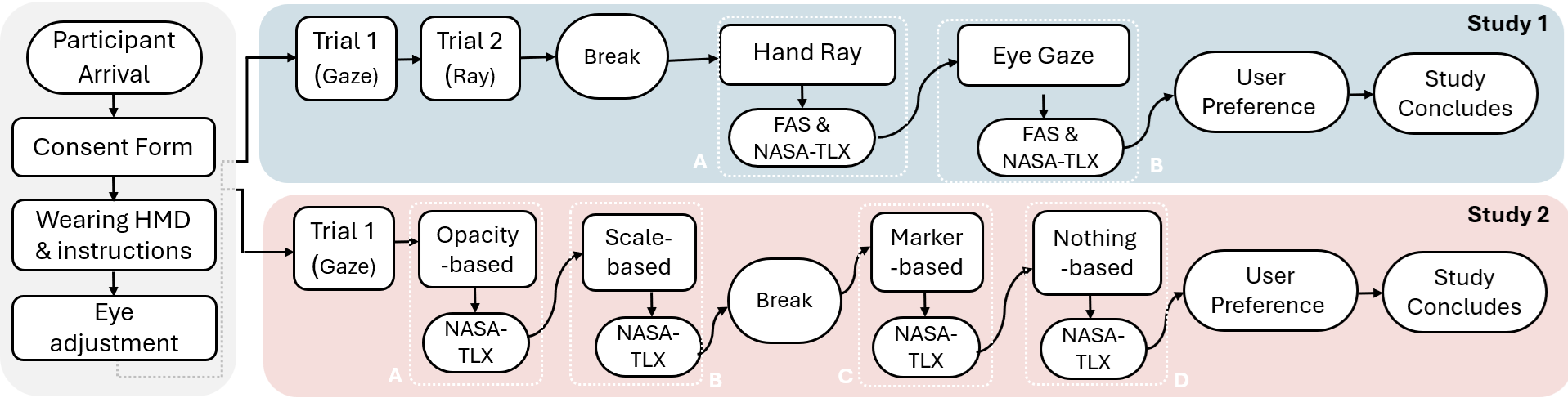}
    \caption{Each participant followed the script shown above throughout one of the above studies. The AR tasks are encompassed in dotted boxes. The order of these tasks was randomized for different participants using a Latin square design in each study.}
    \label{fig:experiment-schedule}
\end{figure*}

Study 2 followed a similar structure. Each task contained five questions, and after each task, participants completed the workload questionnaire. After finishing study 2, they completed a customized user preference questionnaire on the four annotation presentation methods. Each study lasted less than 60 minutes. Figure~\ref{fig:experiment-schedule} presents an overview of the experiment procedure. 

%Note that the order of the tasks shown with the surrounding dotted boxes in Figure~\ref{fig:experiment-schedule} varies for different participants to prevent the ordering effect and bias.

% \begin{table}[]
%     % \footnotesize
%     \centering
%       \caption{Latin square for determining the order of tasks participants was presented with. Tasks are labeled as A, B, C, D, E. \label{tab:latin-square-order-of-tasks}}
%           \begin{tabular}{|l|l|l|l|l|}
%             \hline
%             \textbf{Task 1}               & \textbf{Task 2}  & \textbf{Task 3}  & \textbf{Task 4} & \textbf{Task 5}       \\ \hline
%                  A   & B   & C   & D & E                    
%                  \\ \hline
%                   A  & B  & D  & E & C                    
%                    \\ \hline
%                    A   & B   & E   & C & D 
%                    \\ \hline
%                      C& D & E & B & A  
%                      \\ \hline
%                        D   & E & C & B   & A  
%                      \\ \hline
%                       E  & C & D & B  & A 
%                     \\ \hline
%         \end{tabular}

% \end{table}

\section{Measurements}

Both studies collected objective data including task completion times and the counts of annotations activated via hovering interactions. Regarding subjective data, study 1 employed the Fatigue Assessment Scale (FAS)~\cite{de2004measuring} and the NASA Task Load Index (NASA-TLX)~\cite{hoonakker2011measuring}. Additionally, at the conclusion of Study 1, participants completed a customized comparison questionnaire. In study 2, participants completed the NASA-TLX questionnaire and a customized user preference questionnaire.

\subsection{Objective Measurements}

User data was recorded throughout all the tasks during the experiment. 
We designed scripts to automatically collect participants’ task completion times, their responses to the experiment's questions, and information about the exact times and number of activations for annotations. The completion time for each question was measured from the moment the question appeared until the participant submitted their response. The total time taken to answer all five questions in a task was considered as the task completion time. We also calculated the accuracy of participants’ responses; however, since the questions were designed such that all necessary information was visible in the AR environment and did not require prior knowledge, most participants answered all the questions correctly. Therefore, we do not report the accuracy results. Lastly, we counted the number of activated annotations in each task to compare interaction behavior across different presentation conditions. These activation counts could convey important information regarding the ease of finding the response or inadvertent number of annotation activations.

\subsection{Subjective Measurements}

The FAS is a standardized self-report questionnaire that measures fatigue severity across physical and mental domains~\cite{de2004measuring}. It contains ten items rated on a 5-point Likert scale (1 = not at all, 5 = extremely), with higher scores indicating greater perceived fatigue. To report the final scores, the negatively worded items reverse-scored to maintain consistency. 

The NASA-TLX is a standardized subjective workload assessment tool composed of six dimensions (mental demand, physical demand, temporal demand, performance, effort, and frustration). Each dimension is rated using a 20-Likert-type scale ranging typically from ``very low'' to ``very high'', resulting in an aggregated workload score. Higher total scores indicate higher perceived workload. 

In addition, a customized questionnaire was designed for study 2 to compare annotation presentation methods by capturing user preferences in terms of ease of understanding, ability to maintain focus, level of distraction, perceived performance improvement, confidence in use, and overall intuitiveness. Participants were asked to respond with the annotation presentations they felt best matched each criterion, and they were allowed to select multiple options. Participants were also asked to rank the annotation presentations from 1 to 4, with 1 and 4 indicating their most preferred and least preferred options, respectively. Additionally, they were asked to describe the features they liked or disliked about each visualization (see \S~\ref{subsec:res_study2} for details). 

%The distribution of participants' responses to the FAS and UEQ is shown in Fig~\ref{Figure:SUS3Groups} and~\ref{Figure:UE3Groups}, respectively, where different colors represent scores from 1 to 5. In the figure, red indicates a score of 1 (strong disagreement), while green represents a score of 5 (strong agreement). A greener color suggests a higher positive evaluation of the system, whereas red represents lower ratings, indicating disagreement.

\begin{figure}[t!]
\centering
    \includegraphics[width=.35\linewidth]{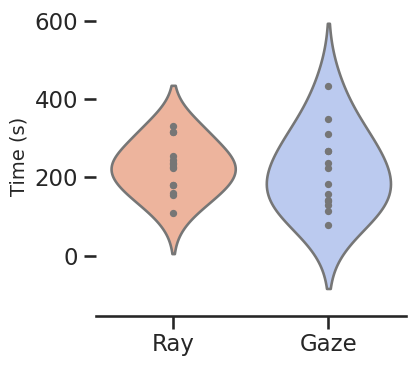}
        \hspace{1.5cm}
    \includegraphics[width=.35\linewidth]{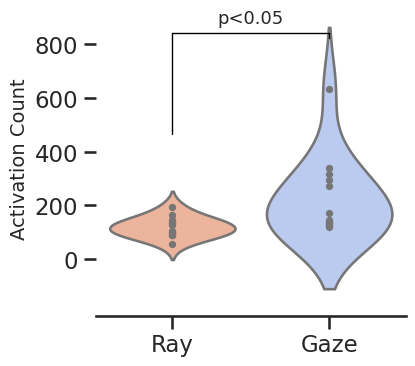}  
\caption{
Comparison of task completion time between the two interaction modalities in Study 1 (left), and comparison of activation count between the two interaction modalities (right). The difference was only statistically significant for task completion times ($p=0.004$).} 
\label{fig:gaze_ray_time_count}
\end{figure}

\section{Results}

In this section, we will elaborate on the results of our experiments and their analysis. We start with the results of study 1 (\S~\ref{subsec:res_study1}) and continue with our empirical observations in study 2 (\S~\ref{subsec:res_study2}).

\subsection{Study 1}
\label{subsec:res_study1}

\paragraph{\textbf{Task completion time}} We evaluated the task completion times for each task of study 1 for all the participants and compared the distribution between the two tasks in Figure~\ref{fig:gaze_ray_time_count} (left). As the violin plots show the average task completion time is slightly larger for hand-ray hovering task; however, this difference did not reach statistical significance in a t-test, and thus $H_1$ was not supported.

\paragraph{\textbf{Activation count}} We also presented the comparison of the activation counts which is presented in Figure~\ref{fig:gaze_ray_time_count} (right). As the violin plots show, participants produced significantly fewer activations with ray interaction technique compared to gaze. This indicates that hand ray offered greater control and reduced unintended activations, whereas gaze hovering often led to accidental triggers. The difference was statistically significant ($p < 0.05$), and therefore $H_2$ was supported.

\begin{figure*}[t!]
    \centering
    \includegraphics[width=.34\linewidth]{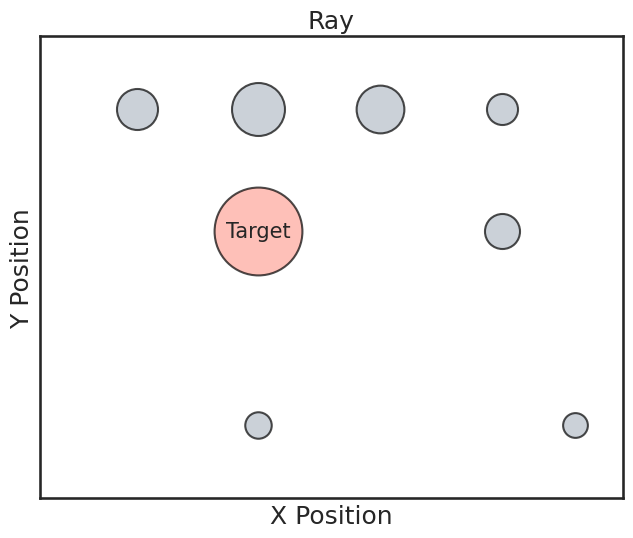}
    \hspace{1.5cm}
    \includegraphics[width=.34\linewidth]{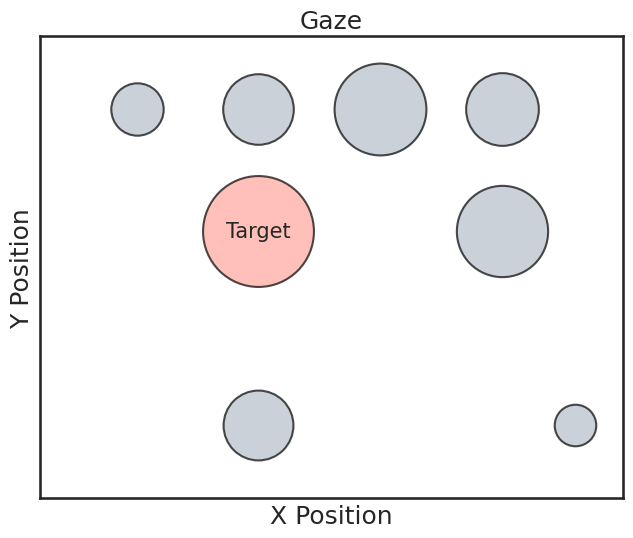}  
\caption{
Comparison of the aggregate activation counts for the target annotation (shown in red) and its nearby annotations (shown in gray), in the questions designed for this analysis. As the figure shows, not only are the activation counts overall larger in Gaze, but they are also relatively larger for annotations that are farther away from the target.}
\label{fig:activation_q3_ray_gaze}
\end{figure*}

For a more detailed and controlled comparison of activation counts, we examined the question in which the target annotation and its neighboring annotations were positioned similarly. This controlled setup allowed us to directly compare how the two interaction modalities influenced activation precision. Figure~\ref{fig:activation_q3_ray_gaze} compares the aggregate activation counts, across all participants, for the target annotation (highlighted in red) and their surrounding annotations (shown in gray) across the two interaction modalities. In the hand-ray condition, activations were concentrated primarily on the target, with relatively few activations occurring on nearby annotations. In contrast, the gaze condition resulted in a higher overall number of activations, not only on the target itself but also noticeably on the neighboring annotations, even on those positioned farther away. This pattern suggests that gaze hovering introduced greater instability and imprecision, leading to more frequent accidental activations outside the intended target. Still, further research is needed to support this observations.

\begin{figure}[t]
    \centering
    \includegraphics[width=0.65\linewidth]{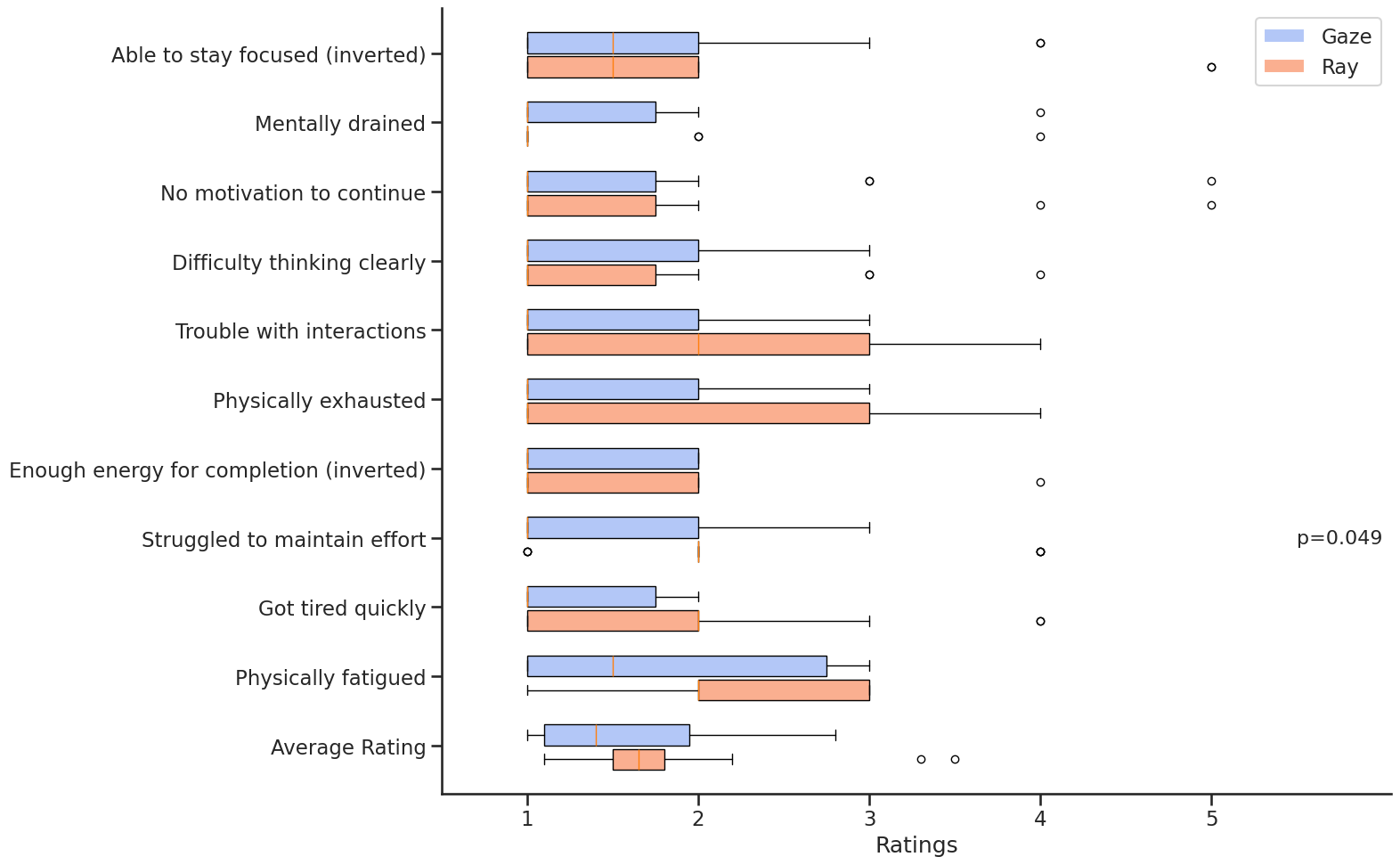}
    \caption{Comparison of FAS scores between the two interaction modalities. The Wilcoxon test is used to derive the significant of the differences. For the significant ones ($p \le 0.05$), the p-values is represented in the plot.}
\label{fig:fas_gaze_ray}
\end{figure}

\paragraph{\textbf{Fatigue Assessment Scale (FAS)}} Figure~\ref{fig:fas_gaze_ray} shows the boxplots of the ratings for each question. We analyzed the significance of differences in the ratings using the Wilcoxon test and reported the p-values for comparisons that were statistically significant ($p \le 0.05$). The results indicate that the hand-ray condition yielded relatively higher ratings on most negative aspects, leading to a slightly worse average score. However, statistical significance was observed only for the comparison of participants’ reported struggle to maintain effort.

%and reduced perceived sustainability during use, despite the hands-free nature of the Gaze modality.

\paragraph{\textbf{ NASA-TLX}} 

\begin{figure}
\centering
\begin{minipage}{.58\textwidth}
  \centering
  \includegraphics[width=.99\linewidth]{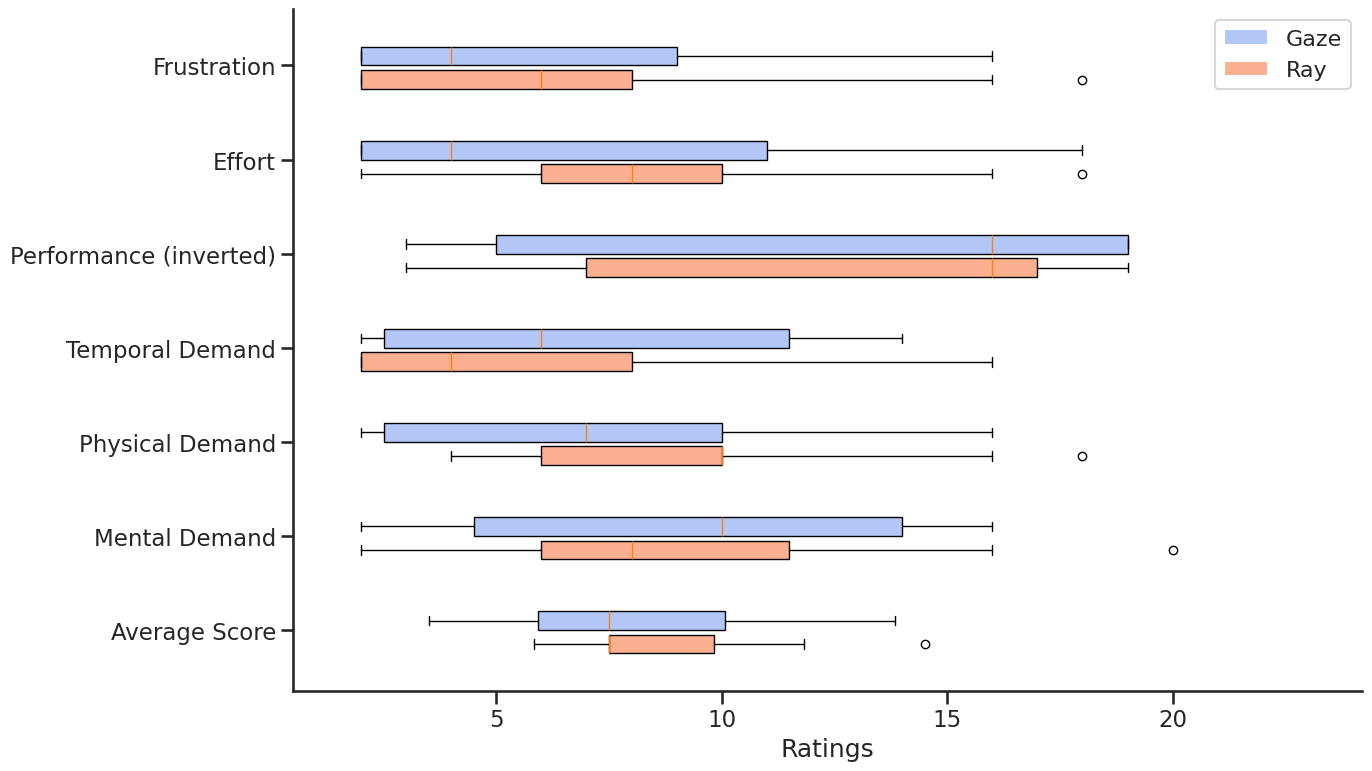}
  \captionof{figure}{Comparison of NASA-TLX scores between the two interaction modalities. The Wilcoxon test is used to derive the significant of the differences. Neither of the the items showed a significant difference ($p \le 0.05$).}
  \label{fig:tlx_task1}
\end{minipage}%
\hspace{0.3cm}
\begin{minipage}{.39\textwidth}
  \centering
  \includegraphics[width=.98\linewidth]{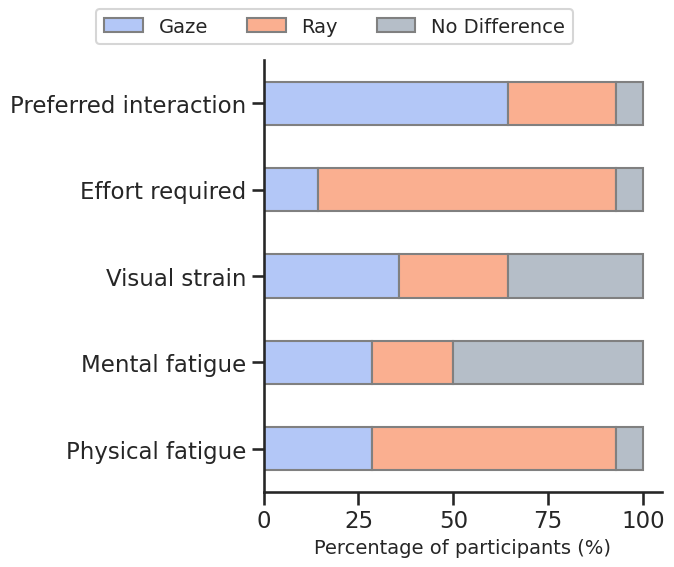}
  \captionof{figure}{Comparison of the percentage of user preference questionnaire between the two interaction modalities across $5$ categories shows on the y-axis.}
  \label{fig:comparison_task1}
\end{minipage}
\end{figure}

As shown in Figure~\ref{fig:tlx_task1}, gaze received higher ratings for mental demand and temporal demand, whereas the hand-ray condition was rated as more physically demanding and effortful. When aggregated into the average workload score, gaze exhibited a slightly lower average workload than hand ray. However, these differences were not statistically significant, and therefore our results did not support $H_4$.

% \begin{figure}[t]
%     \centering
%     \includegraphics[width=0.59\linewidth]{figures/nasa_tlx_task1.png}
%         \includegraphics[width=0.38\linewidth]{figures/task1_comparison_costomized.png}
%     \caption{(Left) Comparison of NASA-TLX scores between the two interaction modalities. The Wilcoxon test is used to derive the significant of the differences. For the significant ones ($p \le 0.05$), the p-values is represented in the plot. (Right) Comparison of the percentage of user preference questionnaire between the two interaction modalities for physical fatigue, mental fatigue, visual strain, effort required, and preferred method.}
% \label{fig:tlx_task1}
% \end{figure}

% \begin{figure}
% \centering
% \begin{subfigure}{.59\textwidth}
%   \centering
%   \includegraphics[width=.4\linewidth]{figures/nasa_tlx_task1.png}
%   \caption{A subfigure}
%   \label{fig:sub1}
% \end{subfigure}%
% \begin{subfigure}{.38\textwidth}
%   \centering
%   \includegraphics[width=.4\linewidth]{figures/task1_comparison_costomized.png}
%   \caption{A subfigure}
%   \label{fig:sub2}
% \end{subfigure}
% \caption{A figure with two subfigures}
% \label{fig:test}
% \end{figure}

\paragraph{\textbf{ User Feedback}}

To capture participants’ opinions on fatigue after experiencing both interaction modalities, we administered a comparison questionnaire. As Figure~\ref{fig:comparison_task1} shows, the results indicate that physical fatigue (64.3\%) and effort required (78.6\%) were most often attributed to Hand ray. Mental fatigue responses were split, with No difference being most common (50.0\%), followed by Eye gaze (28.6\%) and Hand ray (21.4\%). Overall, when we asked participants about their preferred interaction modalities for annotation review task, Eye gaze was preferred by the majority of participants (64.3\%), compared to Hand ray (28.6\%), with only one participant reporting No difference (7.1\%). 

We also asked participants what contributed to their sense of fatigue. Several reported physical strain from hand interaction, including arm fatigue from holding the ray and difficulty in precisely placing it. Others noted the instability of the hand ray and the need to constantly focus on the beam, which increased their workload. On the other hand, while eye gaze required less physical effort, some participants found it more visually tiring because it forced them to fixate their eyes for extended periods. As one participant explained: ``In the ray task, it hurt my arm to keep my hand raised to activate the annotation. In the gaze task, it strained my eyes when annotations I didn’t need were activated, and it was harder than with the ray to select only the annotations I wanted. But in the end, I preferred eye gaze since it didn’t require my hand.''

\subsection{Study 2}
\label{subsec:res_study2}

\begin{figure*}[t!]
    \includegraphics[width=.49\linewidth]{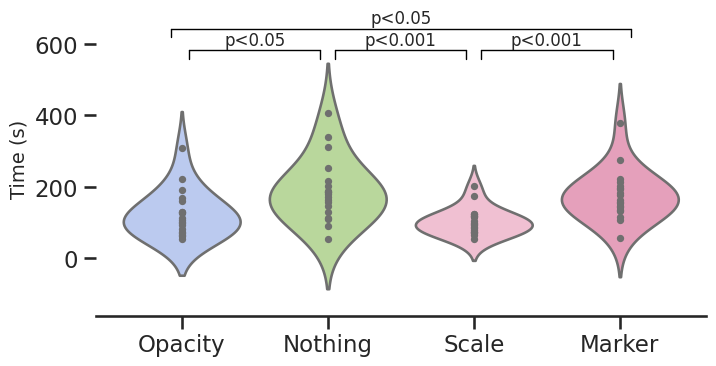}
    \includegraphics[width=.49\linewidth]{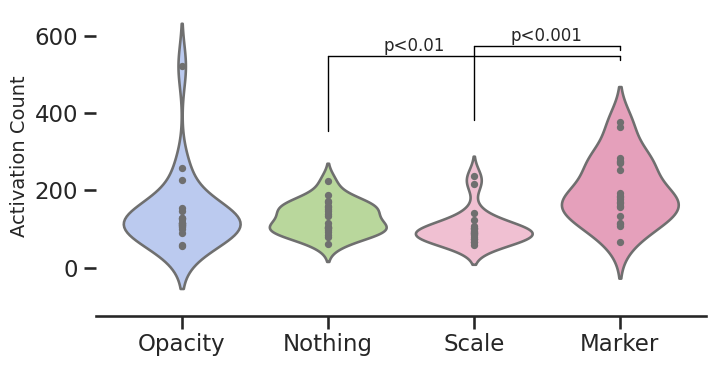}  
\caption{
Comparison of task completion time between the four annotation presentations in Study 2 (left), and comparison of activation count between these presentations (right). The differences among the tasks for both measures are statistically significant using the ANOVA test, with p-values of $0.0002$ and $0.0006$, respectively.} 
\label{fig:task2_time_count}
\end{figure*}

\paragraph{\textbf{Task completion time}} The comparison of task completion times for each task of study 2 is presented in Figure~\ref{fig:task2_time_count} (left). As the violin plots show, the scale-based visualization achieves the lowest average value and its distribution is more condensed around the mean. The highest average value belongs to the Nothing-based presentation. This matches our expectations as it is more difficult for the users to find the annotations in that task where there is no visual clue about the location of annotations.

To evaluate differences in task completion time among the four annotation presentation styles, we conducted a one-way ANOVA test. The analysis revealed a statistically significant difference among these groups ($p = 0.0002$). Given this result, we performed post-hoc pairwise comparisons using paired t-tests to determine which specific pairs of presentation styles differed significantly. The corresponding p-values for each comparison are presented in Table~\ref{tab:ttest}. Statistically significant differences are highlighted in bold, indicating various levels of statistical significance ($p \le 0.05$, $p \le 0.01$, and $p \le 0.001$).

\paragraph{\textbf{Activation count}} We also presented the comparison of the annotation activation counts in Figure~\ref{fig:task2_time_count} (right). As the violin plots show, the Scale-based visualization achieves the lowest average activation count. The largest activation count belongs to the Marker-based task and then the Opacity-based task. This might be due to the fact that the users can see where the annotations are but they need to activate them to find out what type of annotation they are and what content they might represent.

A  one-way ANOVA was conducted to assess differences in activation counts across the four annotation presentations. This analysis also revealed a statistically significant difference ($p=0.0006$). As with task completion time, we performed post-hoc paired Student's t-tests to examine pairwise differences between presentation styles. The results, including p-values for each comparison, are also reported in Table~\ref{tab:ttest}, with bold values indicating various levels of statistical significance ($p \le 0.05$, $p \le 0.01$, and $p \le 0.001$). $H_5$ was supported by significant differences observed in both task completion time and activation count. %$H_5$ was supported by our results. 

%\textbf{Statistical analysis:} To evaluate the differences among the four annotation presentation styles, we conducted a one-way ANOVA. The analysis revealed statistically significant differences between groups for task completion time (p = 0.0009) and activation count (p = 0.0005). Given these significant results, we performed post hoc pairwise comparisons using independent t-tests to examine which specific pairs of presentation styles differed from each other. The resulting p-values for each pairwise comparison are reported in Table~\ref{your_table_label}. Statistically significant differences are indicated in bold, denoting p-values below the standard threshold (typically p < 0.05).

\begin{table*}[th!]
\begin{center}
\begin{small}
\begin{sc}
\resizebox{0.9\textwidth}{!}{
\begin{tabular}{@{} l  c c c c | c c c c @{}}
 \toprule

 % \midrule   
 % & \multicolumn{3}{@{}c}{\textbf{Orig}} & \multicolumn{3}{@{}c}{\textbf{Clip}} \\\addlinespace[0.3em]

& \multicolumn{4}{@{}c}{\textbf{Time}} & \multicolumn{4}{@{}c}{\textbf{Activation Counts}} \\\addlinespace[0.3em]

 \multicolumn{1}{c}{\scriptsize \textbf{}} & 
 \multicolumn{1}{c}{\scriptsize Opacity} & 
 \multicolumn{1}{c}{\scriptsize Nothing} & 
 \multicolumn{1}{c}{\scriptsize Scale} & 
 \multicolumn{1}{c}{\scriptsize Marker} & 
 \multicolumn{1}{c}{\scriptsize Opacity} & 
 \multicolumn{1}{c}{\scriptsize Nothing} & 
 \multicolumn{1}{c}{\scriptsize Scale} & 
 \multicolumn{1}{c}{\scriptsize Marker}
 % &  $N=3$ & $N=9$ & {\footnotesize Improvement}  
 \\\addlinespace[0.3em]

 \cmidrule(r){2-5}
 \cmidrule(r){6-9}

 %  Opacity & -  & $0.03$ & $0.244$ & $0.042$ & 
 
 % -  & $0.557$ & $0.146$ & $0.109$ 
 % \\\addlinespace[0.3em]

 %  Nothing & $0.03$ &  -  & $0.0$ & $0.564$ & 
 
 %   $0.557$ &  -  & $0.1$ & $0.003$ 
 % \\\addlinespace[0.3em]

 %  Scale &  $0.244$ & $0.0$ &  -  & $0.0$ & 
 
 %  $0.146$ & $0.1$ &  -  & $0.0$ 
 % \\\addlinespace[0.3em]

 %   Marker & $0.042$ & $0.564$ & $0.0$ &  -  & 
 
 %  $0.109$ & $0.003$ & $0.0$ &  -   
 % \\\addlinespace[0.3em]

  Opacity & -  & $\bf < 0.05$ & $0.244$ & $\bf < 0.05$ & 
 
 -  & $0.557$ & $0.146$ & $0.109$ 
 \\\addlinespace[0.3em]

  Nothing & $\bf < 0.05$ &  -  & $\bf < 0.001$ & $0.564$ & 
 
   $0.557$ &  -  & $0.1$ & $\bf < 0.01$ 
 \\\addlinespace[0.3em]

  Scale &  $0.244$ & $\bf < 0.001$ &  -  & $\bf < 0.001$ & 
 
  $0.146$ & $0.1$ &  -  & $\bf < 0.001$ 
 \\\addlinespace[0.3em]

   Marker & $\bf < 0.05$ & $0.564$ & $\bf < 0.001$ &  -  & 
 
  $0.109$ & $\bf < 0.01$ & $\bf < 0.001$ &  -   
 \\\addlinespace[0.3em]

 \bottomrule
\end{tabular}
}
\end{sc}
\end{small}
\end{center}
\caption{For task completion time and the activation counts, each value in the table shows the p-value derived from the Students t-test between each pair of tasks. The significant ones are shown in bold.}
\label{tab:ttest}

\end{table*}

\begin{figure}[h]
    \centering
    \includegraphics[width=0.74\linewidth]{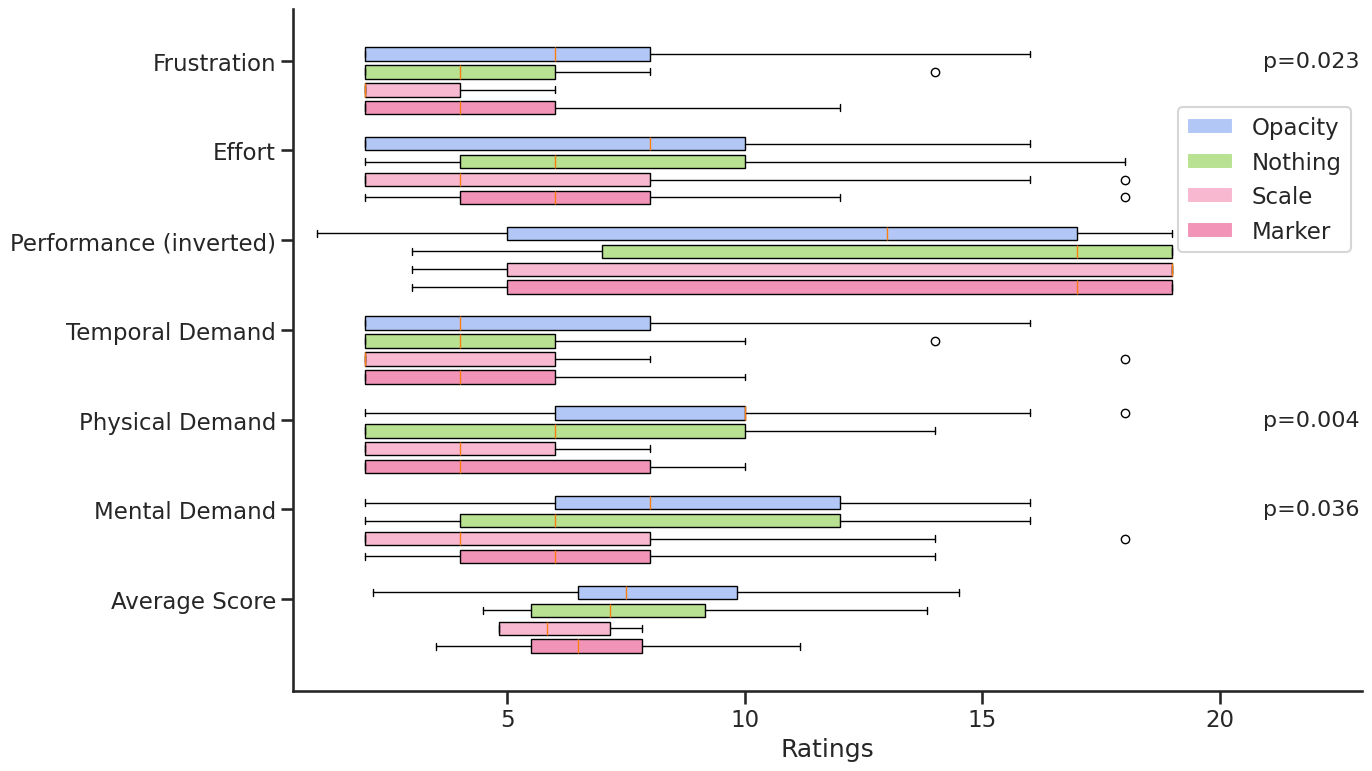}
    \caption{Comparison of NASA-TLX scores between the four presentations. For the group where the Friedman test shows significant difference ($p \le 0.05$), the p-value is represented.}
\label{fig:tlx_task2}
\end{figure}

\paragraph{\textbf{NASA-TLX}}Figure~\ref{fig:tlx_task2} presents the NASA-TLX workload ratings for the four annotation presentation designs across seven dimensions. Overall, the scale-based condition yielded the lowest workload scores, indicating that participants found it to be the least mentally, temporally, and physically demanding, with lower effort and frustration. In contrast, the opacity-based condition consistently showed the highest workload across most categories, including mental and physical demand, effort, and average workload. Marker-based annotations performed moderately well, while the nothing-based condition showed mixed results.

\begin{figure*}[h]
    \includegraphics[width=0.48\linewidth]{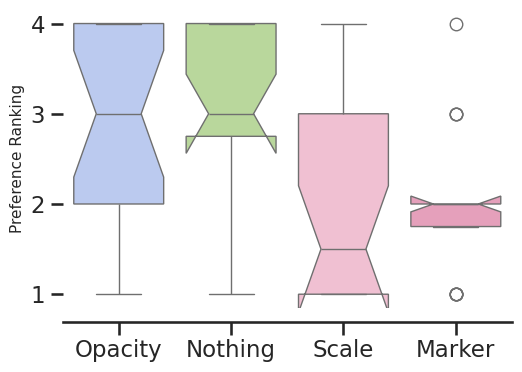}
    \includegraphics[width=.48\linewidth]{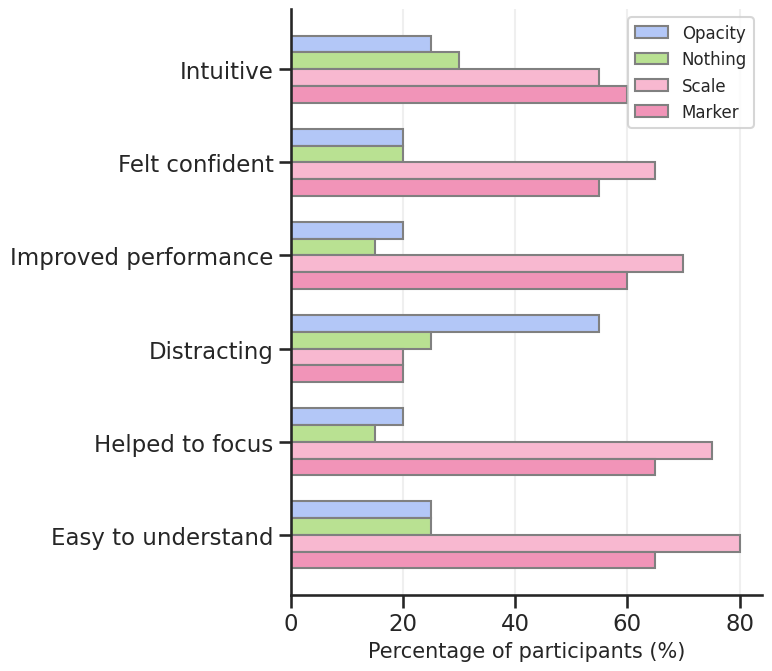}  
\caption{
Comparison of participants preference ranking between the four presentations. The notch shows the median value for each box-plot. The Friedman test on these rankings shows that the differences among these groups is significant ($p=0.003$) (left). comparison of the visualization methods using a customized usability questionnaire covering six categories: intuitiveness, confidence, performance enhancement, distraction, focus, and ease of understanding (right). } 
\label{fig:preference_task2}
\end{figure*}

%These findings suggest that scale-based designs offer the best balance between clarity and usability, while opacity- and nothing-based designs may hinder performance depending on task complexity and user familiarity.
\paragraph{\textbf{Preference rating}} The preference ratings that were collected from the users after completing study 2 was combined to make a comparison for user-based preferences for the visualization types. Figure~\ref{fig:preference_task2} (left) shows the box-plots for these ratings. The median value of the rankings has been shown with a notch. As this figure shows, the Scale-based visualization has the best ranking among the visualizations by a median of $1.5$. The next preferred form of the visualization was the Marker-based one, with a median of $2$. Nothing-based and Opacity-based visualizations were the least preferred visualizations with a median of $3$.  

We also performed the Friedman test to evaluate if the differences in the four groups are significant. We derived a p-value of $0.003$ for this test, which confirms the significant difference in the user-preferences, thereby supporting $H_6$. The highest preference, which is for the Scale-based visualization, matches the results from the task-completion times and activation counts in Figure~\ref{fig:task2_time_count}; however, it is interesting to note that although the Marker-based presentation method was among the worst in terms of task completion time and activation counts, it was the second preferred option by the users, highlighting the subjective nature of user satisfaction and the necessity for collecting user preferences in addition to evaluating the objective measures.

Other than the rankings, we asked the participants to choose among the visualization methods based on six categories: intuitiveness, providing confidence, performance enhancement, being distracting, helping to focus, and being easy to understand. Each user was allowed to choose zero or multiple methods for each category. Figure~\ref{fig:preference_task2} (right) shows the percentage of participants that have chosen each of the visualization methods for each category. Scale-based method consistently outperformed the other conditions, with the highest percentage of participants reporting that it was intuitive, easy to understand, helped them focus, improved performance, and increased confidence. Marker-based was the second most positively rated method, particularly in aiding focus and performance. In contrast, Opacity-based method was frequently rated as distracting, with nearly half of participants identifying it as such, and it received comparatively lower ratings in positive categories.

In addition to the rankings, participants evaluated the visualization methods across six categories: intuitiveness, confidence, performance enhancement, distraction, focus support, and ease of understanding. Participants could select multiple methods for each category. As shown in Figure~\ref{fig:preference_task2} (right), the Scale-based method received the highest endorsement across most dimensions, being frequently described as intuitive, easy to understand, supportive of focus, performance-enhancing, and confidence-building. The Marker-based method ranked second, particularly for being intuitive and easy to understand. By contrast, the Opacity-based method was often considered distracting and was less favored in the positive categories.

%We also performed the Friedman test which is a non-parametric test used when the dependent variable is ordinal (e.g., rankings) and shows if the difference in three or more groups are significant

%These findings suggest that while Gaze may offer hands-free control, it could also introduce greater physical and cognitive demand compared to Ray-based interaction.

\paragraph{\textbf{User feedback:}}
 In the customized questionnaire conducted after the four presentation methods, we asked participants to share their feedback on what they liked and disliked about each of the methods. A thematic analysis was applied to their feedback, and the results are presented in Table~\ref{tab:presentation_thematic_analysis}. Participants appreciated how the zooming feature in the Scale-based method helped them focus and made it easier to see important items. Some also mentioned that the smaller version of each annotation helped guide them (e.g., ``Giving a little knowledge about what is in, before opening it'). However, some participants found the zooming effect distracting (e.g., ``Distraction while focusing on content''). The Marker-based visualization was praised for its clear signs and ease of navigation. As one participant noted, ``Markers made it easy to find and distinguish annotations from the background.'' The Nothing-based condition was seen as simple and not visually distracting by a few participants, but most found it difficult to locate information without any visual cues. The Opacity-based method was the least preferred, with some users saying it felt cluttered (e.g., ``It was cluttering the view, and several icons were overlapping''). We also asked participants to provide examples of contexts in which they would prefer each presentation method. Scale-based was mentioned for tasks such as reading, vision support, museum visits, campus tours, multitasking, component design and digital twins, and quick navigation. Nothing was preferred for driving, gaming, shopping in grocery stores. Marker-based was considered useful for repair and maintenance, multitasking, labeling objects, and handling overlapped annotations. Opacity-based was described as helpful when everything should remain visible in place, but not necessarily as the main focus.

% \begin{table*}[ht]
% \centering
% \caption{Thematic analysis of participant feedback on the four presentation methods (percentages are based on 20 participants).}
% \begin{tabular}{|p{1.5cm}|p{5cm}|p{5cm}|}
% \hline
% \textbf{Modality} & \textbf{Strengths} & \textbf{Weaknesses}  \\
% \hline
% \textbf{Opacity} &
% - Helped finding target (35\%) \newline
% - Helped ignore irrelevant info (15\%)  &
% - Distracting or cluttered (30\%) \newline
% - Confusing (15\%) \\
% \hline
% \textbf{Scale} &
% - Intuitive \& easy to use (20\%) \newline
% - Fast navigation (15\%) \newline
% - Compact view (25\%) \newline
% - Better readability (25\%) &
% - Distraction during content focus (15\%) \\
% \hline
% \textbf{Marker} &
% - Easier navigation (45\%) \newline
% - Improved target finding (15\%) \newline
% - Less distraction/cluttered (10\%) 
% &
% - Hard to find target (20\%) \newline
% - Less informative (10\%) \newline
% - Distraction during content focus (10\%) \\
% \hline
% \textbf{Nothing} &
% - Less distraction (40\%) \newline
% - Intuitive \& easy to use (10\%) &
% - Hard to navigate (40\%) \newline
% - Struggle to find target (20\%) 
% \\
% \hline
% \end{tabular}
% \label{tab:presentation_thematic_analysis}
% \end{table*}

\begin{table*}[t]
\centering
\caption{Thematic analysis of participant feedback on the four presentation methods (percentages are based on 20 participants).}
\label{tab:presentation_thematic_analysis}

% local spacing
{\renewcommand{\arraystretch}{1.2}%
 \setlength{\tabcolsep}{6pt}%

\begin{tabularx}{\textwidth}{@{} l >{\raggedright\arraybackslash}X >{\raggedright\arraybackslash}X @{}}
\toprule
\textbf{Modality} & \textbf{Strengths} & \textbf{Weaknesses} \\
\midrule
\textbf{Opacity} &
-- Helped finding target (35\%) \newline
-- Helped ignore irrelevant info (15\%) &
-- Distracting or cluttered (30\%) \newline
-- Confusing (15\%) \\

\textbf{Scale} &
-- Intuitive \& easy to use (20\%) \newline
-- Fast navigation (15\%) \newline
-- Compact view (25\%) \newline
-- Better readability (25\%) &
-- Distraction during content focus (15\%) \\

\textbf{Marker} &
-- Easier navigation (45\%) \newline
-- Improved target finding (15\%) \newline
-- Less distraction/clutter (10\%) &
-- Hard to find target (20\%) \newline
-- Less informative (10\%) \newline
-- Distraction during content focus (10\%) \\

\textbf{Nothing} &
-- Less distraction (40\%) \newline
-- Intuitive \& easy to use (10\%) &
-- Hard to navigate (40\%) \newline
-- Struggle to find target (20\%) \\
\bottomrule
\end{tabularx}%
}
\end{table*}

%Overall, participants' feedback shows that visual cues can improve interaction and reduce mental effort in complex AR environments.

\section{Discussion}

\paragraph{RQ1. Which interaction modality, across AR and VR, provided higher performance and user preference?} Our findings suggest that, on average, eye-gaze hovering enabled faster task completion and reduced fatigue and workload compared to hand-ray hovering. Although the differences were mostly not statistically significant, the preference post-questionnaire showed a clear win for eye-gaze hovering in terms of the required effort and physical fatigue felt by the users, as well as overall preference. This trend is consistent with prior works. Tadeja et al.\cite{tadeja2024using} reported that eye gaze outperformed hand ray in speed within a multi-interaction system using pinch gestures for selection, and Lee et al.\cite{lee2025experimental} found that head gaze produced lower workload compared to hand ray. 

However, this efficiency came with a trade-off—a significant higher rate of unintended activations when using gaze hovering. The imprecision can be attributed to the natural instability of eye movements and involuntary fixations, a well-known issue often referred to as the ``Midas Touch'' problem. In contrast, hand-ray interaction required more deliberate pointing, which reduced accidental triggers and offered greater control over annotation selection. Overall, these results highlight a balance between speed and reduced physical effort (gaze) versus precision and intentionality (hand-ray).
~\paragraph{RQ2. Among the four annotation retrieval presentation methods, which one yields better performance and user preference?} The results from Study 2 showed that Scale-based presentation consistently outperformed other designs, with shorter task completion times, fewer activation counts, and higher preference ratings. One explanation is that Scale-based presentation provided both compact overviews and expanded visibility when needed, reducing visual clutter while supporting readability. Participants benefited from being able to quickly scan small representations, locate the target object, and then focus their attention on it. By contrast, the Marker-based presentation, while less efficient, was valued for its clear spatial cues. This highlights a distinction between performance efficiency and perceived navigational support. The Opacity-based condition was often described as distracting, likely because changing transparency increased visual noise and competed with background content. Similarly, the Nothing-based condition, while eliminating clutter, made searching more difficult since users lacked guidance on where annotations were located. Both results point to the importance of providing just enough visual cues without overwhelming the display.

\subsection{Design Implications}
\label{subsec:design_implication}

Based on our empirical findings, we propose the following design recommendations for building effective AR annotation review systems:

 \begin{itemize}
     \item Prioritize eye-gaze hovering for lightweight review: eye-gaze hovering seems a rapid, hands-free, and user-friendly interaction modality in annotation review systems compared to hand-ray. This is especially essential in many real-world scenarios where users' hands are occupied.
   
    % \item  Use hand-ray hovering for precise Control in tasks hand-ray interaction is better suited for tasks that require intentionality and precision. 

    \item Implement scale-based annotation presentation as the default option. This presentation method achieves a good balance in the trade-off of avoiding visual clutter and ease of access to the augmented information by dynamically adjusting the annotation size. 

    \item Avoid over-reliance on minimalist approaches: while nothing-based designs can help reduce visual clutter, they may leave users disoriented due to the lack of guidance. However, this approach can be beneficial in scenarios where annotation locations are repetitive and users are already familiar with the environment, allowing them to rely on memory and spatial awareness rather than visual cues.

    \item Support personalization and adaptation: Although this study favored gaze-based interaction for review systems and scale-based visualization for annotation presentation, individual preferences varied, indicating that user needs and task demands are not uniform. This variability highlights the importance of supporting customization in future systems and allowing users to choose their preferred interaction modality and presentation style. Additionally, adaptive interfaces that respond to user behavior and task context could further improve performance and overall user satisfaction.
 \end{itemize}

% \begin{figure}[t]
%     \centering
%     \includegraphics[width=0.7\linewidth]{figures/preference_ranking.png}
%     \caption{Comparison of participants preference ranking between the four presentations. The notch shows the median value for each box-plot. The Friedman test on these rankings shows that the differences among these groups is significant ($p=0.001$).}
% \label{fig:preference_task2}
% \end{figure}

\section{Limitations and Future Work}
\label{sec:future}

While user preferences and performance were assessed, the study did not analyze long-term usability, learning effects, or retention over time. Future work should explore how users adapt to these interaction techniques and presentation styles during prolonged use. Also, in this work, we only studied the differences when users perform simple tasks; future work should better focus on the effectiveness of these methods in performing more complex tasks where users need to retrieve multiple pieces of information to acquire a higher-level knowledge before making a decision. Another interesting problem to study is to evaluate the optimum amount of scaling when the scale-based presentation method is used. Future work can even focus on dynamically tuning these scales depending on the surrounding environment and the visual clutter in the vicinity of the annotations. Another future direction is to combine various representation methods in the same environment depending on their context and neighboring objects.
 
\section{Conclusion}
\label{sec:conclusion}

This work investigated the challenges of annotation retrieval in AR and explored how different interaction modalities and presentation designs impact user performance, workload, fatigue, and preference. Through two controlled user studies, we compared eye-gaze and hand-ray hovering as input methods and evaluated four annotation presentation strategies: Opacity-based, Scale-based, Marker-based, and Nothing-based.

Our findings show that eye-gaze hovering enables faster task completion with less fatigue, making it a promising option for hands-free review tasks. However, it also introduces challenges related to unintentional activation. Hand-ray hovering, while more deliberate and precise, led to increased task time and higher fatigue levels. Among the presentation techniques, scale-based presentation were favored as the most effective and preferred, providing clear visibility with minimal clutter. Marker-based designs also received positive user feedback despite being less efficient, while Opacity-based and Nothing-based approaches were the least favored approaches among participants.

%%
%% The acknowledgments section is defined using the "acks" environment
%% (and NOT an unnumbered section). This ensures the proper
%% identification of the section in the article metadata, and the
%% consistent spelling of the heading.
%\begin{acks}
%[REMOVED FOR SUBMISSION]
%\end{acks}

%%
%% The next two lines define the bibliography style to be used, and
%% the bibliography file.
\bibliographystyle{ACM-Reference-Format}
\bibliography{AnnotationRetreival}

%%
%% If your work has an appendix, this is the place to put it.
\appendix

%\section{Research Methods}

%\subsection{Part One}

%Lorem ipsum dolor sit amet, consectetur adipiscing elit. Morbi

%\subsection{Part Two}

%Et

%\section{Online Resources}

%Nam id fermentum dui. Suspendisse sagittis tortor a nulla mollis, in

\end{document}